\documentclass[twocolumn]{aastex631}

\hypersetup{linkcolor=red,citecolor=blue,filecolor=cyan,urlcolor=magenta}

\shorttitle{N-S Asymmetry of Solar Activity}
\shortauthors{L.\,L.~Kitchatinov and A.\,I.~Khlystova}

\def\bl{\par\vskip 12pt\noindent}
\renewcommand{\vec}[1]{\mbox{\boldmath $#1$}}

\begin{document}

\title{Dynamo Model for North-South Asymmetry of Solar Activity}


\author{Leonid Kitchatinov}
%
\author{Anna Khlystova}
\affiliation{Institute of Solar-Terrestrial Physics SB RAS,
Lermontov Str. 126A, 664033, Irkutsk, Russia}

\begin{abstract}
Observations reveal a relatively small but statistically significant North-South (NS) asymmetry in sunspot activity varying on a time scale of several solar cycles. This paper proposes a dynamo model for the phenomenon of long-term NS asymmetry. The model separates dynamo equations for magnetic fields of dipolar and quadrupolar equatorial parity. The NS asymmetry results from the superposition of dipolar and quadrupolar fields. Model computations confirm the formerly proposed excitation of the quadrupolar dynamo mode by a dominant dipolar mode mediated by the equator-symmetric fluctuations in the $\alpha$-effect as a mechanism for the long-term NS asymmetry. An analytically solvable example of oscillations excited by short-term random forcing is given to justify the numerical result of NS asymmetry coherent on a time scale of several (about 6 in the present model) solar cycles resulting from random variations in the $\alpha$-effect on a time scale of one solar rotation. The model computations show the phase locking phenomenon of dipolar and quadrupolar fields oscillating predominantly in phase (northern type asymmetry) or in antiphase (southern type asymmetry) with relatively short irregular transitions between these two states. Large asymmetry in the simulated Grand minima is found and explained by weak magnetic quenching of the $\alpha$-effect during the minima. The possibility of polar field asymmetry in activity minima as a precursor of sunspot asymmetry in the following activity cycles is discussed based on the dynamo model and observations.
\end{abstract}
\section{Introduction}
The emergence and evolutions of spots on the Sun have a substantial randomness. Positions and lifetimes of sunspots allow statistical regularities only \citep{H96,S03}. The formation of an active region in, say, the northern hemisphere does not normally have a counterpart in the southern hemisphere. The statistics of sunspots and other types of solar magnetic activity are therefore uneven about the solar equator \citep[Sect. 4.11 in][and references therein]{H15}.

The NS asymmetry of solar activity is not completely random however. It has a regular long-term component in line with its (dominant) random part. Long periods of NS asymmetry of a certain type have been known since \citet{S1889} and \citet{M04}.
\citet{RNR93} found that almost all sunspots observed during the Maunder minimum were in the southern hemisphere. \citet{OB94} reported a secular wave-type trend in the NS asymmetry of sunspot activity for 1874-1989 with northern and southern hemispheres alternately dominant for several activity cycles. \citet{BO11} generally confirmed this trend and extended it to the epoch of 1821-2009. \citet{Nea19} found that the statistics of sunspot groups for 7 of 12 solar cycles 12 to 23 are not compatible with the equal probability of individual group emergence in either hemisphere, and the neighboring cycles tend to have asymmetry of the same sense. The presence of statistically significant long-term hemispheric asymmetry in solar activity makes it a subject for dynamo theory.

Solar dynamo models allow global modes of dipolar (odd) and quadrupolar (even) equatorial symmetry \citep{KR80}. Modes of either symmetry type are symmetric in terms of magnetic energy density. The hemispheric asymmetry in magnetic energy demands a superposition of dipolar and quadrupolar fields. \citet{SC18} have shown that the superposition of dipolar and quadrupolar dynamo modes gives long-term oscillations in NS asymmetry with sum and beat frequencies of the dynamo modes.

The parameters of solar dynamo models are known to be close to a certain type of equatorial symmetry (e.g., differential rotation is nearly symmetric while the meridional flow and the mean $\alpha$-effect are antisymmetric about the equator). Models with parameters strictly obeying their corresponding symmetry produce magnetic fields of a certain equatorial parity. Dynamo models for NS asymmetry have to include deviation of their parameters from their corresponding symmetry. Two known possibilities for the deviations are 1) the nonlinear dynamo effects, and 2) random fluctuations in dynamo parameters.

Parameters of nonlinear dynamo models depend on the magnetic field. They deviate from an equatorial symmetry if the field is asymmetric. Sufficiently strong nonlinearities can result in a self-sustained hemispheric asymmetry \citep{SNR94,Tob97,WT16,SK17}. However, observational gyrochronology shows that solar rotation is close to the rate where other stars of comparable mass cease their magnetic activity \citep{vSea16,MvS17}. The solar dynamo is therefore only slightly supercritical and weakly nonlinear \citep{CS17,KN17}.

Fluctuations in dynamo parameters can be responsible for different kinds of variability in solar activity cycles \citep[see Chapter~7 in][and references therein]{Cha20}. Asynchronous fluctuations in the southern and northern hemispheres can cause asymmetry in magnetic activity \citep{SC18,Kea18,NLC19,Mea20}.

\citet{Nea19} suggested that NS asymmetry in solar activity results from equator-symmetric fluctuations in the $\alpha$-effect that transform the dominant dipolar dynamo mode into the quadrupolar mode.
Their dynamo model produced long-term asymmetry that varied aperiodically with a coherence time of about four activity cycles. A counterintuitive feature of the model however is that the long-term asymmetry resulted from fluctuations with a short correlation time of one solar rotation. It is unclear whether such short fluctuations with zero mean can result in coherent quadrupolar oscillations with a period about one hundred times longer. In this paper, we test and confirm this possibility by explicit computations of dipolar and quadrupolar field dynamics in a dynamo model with fluctuating $\alpha$-effect. The dynamo model of \citet{KMN18} is modified to allow splitting the dynamo equations into separate equations for magnetic fields of dipolar and quadrupolar parity. Computed statistics of about one thousand dynamo cycles clearly show that the excitation of the quadrupolar mode by the dominant dipolar mode mediated by short-term equator-symmetric fluctuations in the $\alpha$-effect can produce long-term hemispheric asymmetry in magnetic energy.
We also show that the excitation of quasi-periodic oscillations by random
forcing, the correlation time of which is short compared with the oscillation
period, is not an exceptional property of dynamos by giving an
analytically solvable example for such a process.

Our dynamo model is described in Section 2. Section 3 gives an example of long-term oscillations excited by short-term random forcing and introduces the method of oscillation analysis that is afterwards applied to the results of dynamo computations. Sect.\,4 presents and discusses the results of dynamo modeling. Section 5 summarizes the results and concludes.
\section{Dynamo model}
Our dynamo model is a particular version of the so-called flux-transport models initiated by \citet{Dur95} and \citet{CSD95}. This name reflects the significance of the meridional flow for latitudinal migration of the magnetic fields. The flux-transport models with nonlocal $\alpha$-effect of the Babcock-Leighton (BL) type are known to agree closely with solar observations \citep{Jea13}.

The model of this paper differs from the model of \citet{KN17,KN17AL} in its formulation of the $\alpha$-effect and in its separate treatment of dipolar and quadrupolar parts of the magnetic field. We therefore describe these differences in all necessary detail and outline briefly the rest of the model design.
\subsection{Fluctuating $\alpha$-effect}
Our mean-field dynamo model solves numerically 2D dynamo equations for the magnetic field in the spherical shell of the convection zone. The cylinder symmetric magnetic field $\vec{B}$  can be written as
\begin{equation}
    \vec{B}= \vec{e}_\phi B(r,\theta)
    + \vec{\nabla}\times\left(\vec{e}_\phi\frac{A(r,\theta)}{r\sin\theta}\right),
    \label{1}
\end{equation}
where the standard spherical coordinates are used ($\theta$ is the colatitude), $B$ is the toroidal field, $A$ is the poloidal field potential equal to the magnetic flux through a circle of given $r$ and $\theta$, and $\vec{e}_\phi$ is the azimuthal unit vector. The poloidal field obeys the dynamo equation
\begin{equation}
    \frac{\partial A}{\partial t} = -\vec{V}^\mathrm{m}\cdot\vec{\nabla} A + r\sin\theta\ \cal{E}_\phi,
    \label{2}
\end{equation}
where $\vec{V}^\mathrm{m}$ is the meridional flow velocity, and $\vec{\cal{E}} = \langle\vec{u}\times\vec{b}\rangle$ is the mean electromotive force resulting from correlated  fluctuations of the velocity ($\vec{u}$) and the magnetic field ($\vec{b}$) \citep{KR80}.

The azimuthal component $\cal{E}_\phi^\alpha$ of the electromotive force responsible for the $\alpha$-effect is specified as follows:
\begin{equation}
    {\cal E}_\phi^\alpha = \alpha\ \frac{\sin^{n_\alpha}\theta\ \phi_\alpha(r)}{1 + E_\mathrm{m}/E_0}
    \left( \cos\theta + \sigma f(\theta,t)\right) B(r_\mathrm{i},\theta) ,
    \label{3}
\end{equation}
where $E_\mathrm{m}$ is the total magnetic energy in the convection zone and $E_0 = 2\times 10^{38}$\,erg is the model parameter. The $\alpha$-effect of this equation is nonlocal in radius. It generates the poloidal field near the surface from the bottom toroidal field; $r_\mathrm{i}$ is the inner boundary of the convection zone and the function $\phi_\alpha(r)$ peaks near the surface \citep[see Equation\,(18) in][]{KN17}.
The nonlocal $\alpha$-effect is intended to represent the BL mechanism of the poloidal field production near the surface by the magnetic flux-tube rise from the bottom.
The BL mechanism is related to finite tilt of bipolar magnetic regions relative to the line of latitude, the tilt presumably results from the Coriolis force action on raising flux-tubes. Dispersal of magnetic fields of decaying bipolar regions over the solar surface contributes to the global poloidal field \citep[see Chapter~3 in][for more details]{Cha20}.
The relatively large value of $n_\alpha = 7$ in Eq.\,(\ref{3}) of the model reflects the sunspot emergence near the solar equator \citep{Kit20}.
The substantial scatter in the tilt of bipolar magnetic regions around its mean value \citep{H96,JCS14} indicates large fluctuations in the $\alpha$-effect. In Equation\,(\ref{3}), $\sigma = 2.7$ is the relative amplitude of the fluctuations \citep{OCK13}, and $f(\theta, t)$ is the random function of time,
\begin{equation}
    f(\theta, t) = \cos\theta\ S^\mathrm{a}(t)/\sqrt{2} + 0.1 S^\mathrm{s}(t) ,
    \label{4}
\end{equation}
including its symmetric ($S^\mathrm{s}$) and asymmetric ($S^\mathrm{a}$) parts about the equator. The numerical factors in Equation\,(\ref{4}) were introduced in an attempt to equalize the amplitudes of equator-symmetric and antisymmetric fluctuations.

We model the random functions by solving numerically the equation system
\begin{eqnarray}
    \frac{\mathrm{d}S^\mathrm{a}}{\mathrm{d}t} &=& \frac{3}{\tau}\left(S^\mathrm{a} - S_1\right) ,
    \nonumber \\
    \frac{\mathrm{d}S_1}{\mathrm{d}t} &=& \frac{3}{\tau}\left(S_1 - S_2\right) ,
    \nonumber\\
    \frac{\mathrm{d}S_2}{\mathrm{d}t} &=& \frac{3}{\tau}\left(S_2 - \hat{g}\sqrt{\frac{2\tau}{\Delta t}}\right),
    \label{5}
\end{eqnarray}
and the same but independent equations for $S^\mathrm{s}(t)$, in line with the dynamo equations. In this equation, $\hat{g}$ is a normally distributed random number with {\sl rms} value equal one and renewed at each time step of numerical integration, $\tau$ is the correlation time (equal to 25.4~days of one solar rotation in this paper), and $\Delta t$ is the numerical time step. The equations (\ref{5}) give a smoothly varying random function of time with $\langle S^2(t)\rangle \simeq 1$ (see the discussion by \citet{KMN18} of their Equation (3) for more details).

The denominator in Equation\,(\ref{3}) accounts for magnetic quenching of the $\alpha$-effect. The global quenching with total magnetic energy permits separation of dynamo equations for dipolar and quadrupolar parity parts of the magnetic field that would be not possible with traditional quenching by energy density.
\subsection{Dynamo equations for dipolar and quadrupolar fields}
Any 2D magnetic field of Equation\,(\ref{1}) can be split in its dipolar,
\begin{eqnarray}
    B^\mathrm{d}(r,\theta) &=& \left( B(r,\theta) - B(r,\pi - \theta)\right)/2,
    \nonumber\\
    A^\mathrm{d}(r,\theta) &=& \left( A(r,\theta) + A(r,\pi - \theta)\right)/2,
    \label{6}
\end{eqnarray}
and quadrupolar,
\begin{eqnarray}
    B^\mathrm{q}(r,\theta) &=& \left( B(r,\theta) + B(r,\pi - \theta)\right)/2,
    \nonumber\\
    A^\mathrm{q}(r,\theta) &=& \left( A(r,\theta) - A(r,\pi - \theta)\right)/2,
    \label{7}
\end{eqnarray}
parts:
\begin{equation}
    B = B^\mathrm{d} + B^\mathrm{q},\ \ A = A^\mathrm{d} + A^\mathrm{q} .
    \label{8}
\end{equation}
The dipolar (quadrupolar) fields are antisymmetric (symmetric) relative to the mirror-reflection about the equatorial plane.

Before formulating equations for the fields of these two parities, it should be noted that the mean electromotive force in Equation\,(\ref{2}) includes the eddy diffusion and diamagnetic pumping terms in line with the $\alpha$-effect:
\begin{equation}
    {\vec{\cal E}} = {\vec{\cal E}}^\mathrm{dif} + {\vec{\cal E}}^\mathrm{dia} + {\vec{\cal E}}^\alpha ,
    \label{9}
\end{equation}
where
\begin{equation}
    {\vec{\cal E}}^\mathrm{dif} = -\eta{\vec\nabla}\times{\vec B}
    - \eta_\|{\vec e}\times\left({\vec e}\cdot{\vec\nabla}\right){\vec B}
    \label{10}
\end{equation}
is the diffusive part and
\begin{equation}
    {\vec{\cal E}}^\mathrm{dia} = -({\vec\nabla}\tilde{\eta})\times{\vec B}
    + ({\vec\nabla}\eta_\|)\times{\vec e}\left({\vec e}\cdot{\vec B}\right)
    \label{11}
\end{equation}
accounts for the diamagnetic pumping. In these equations, ${\vec e}$ is the unit vector along the rotation axis, and all details about the radius-dependent coefficients $\eta(r),\ \eta_\|(r)$ and $\tilde{\eta}(r)$ are given in \citet{KN17}.

With these equations for the electromotive force, Equation\,(\ref{2}) leads to the following equation for the dipolar part of the poloidal field:
\begin{eqnarray}
    \frac{\partial A^\mathrm{d}}{\partial t}
    &=& - \left(V^\mathrm{m}_r + V^\mathrm{dia}_r\right)\frac{\partial A^\mathrm{d}}{\partial r}
    - \left(V^\mathrm{m}_\theta + V^\mathrm{dia}_\theta\right)\frac{1}{r}
    \frac{\partial A^\mathrm{d}}{\partial \theta}
    \nonumber \\
    &+& \eta\frac{\partial^2A^\mathrm{d}}{\partial r^2}
    + \frac{\eta}{r^2}\sin\theta\frac{\partial}{\partial\theta}\frac{1}{\sin\theta}
    \frac{\partial A^\mathrm{d}}{\partial\theta} + \eta_\|\frac{\partial^2A^\mathrm{d}}{\partial z^2}
    \nonumber \\
    &+& r \alpha\ \frac{\sin^{n_\alpha + 1}\theta\ \phi_\alpha(r)}{1 + E_\mathrm{m}/E_0}
    \bigg( \cos\theta B^\mathrm{d}(r_\mathrm{i},\theta)
    \nonumber \\
    &+&\frac{\sigma}{\sqrt{2}}S^\mathrm{a}(t)\cos\theta B^\mathrm{d}
    + 0.1\sigma S^\mathrm{s}(t) B^\mathrm{q}(r_\mathrm{i},\theta)\bigg),
    \label{12}
\end{eqnarray}
where $\partial/\partial z = \cos\theta\ \partial/\partial r - r^{-1}\sin\theta\ \partial/\partial\theta$ is the spatial derivative along the rotation axis, and two components of the diamagnetic pumping velocity read
\begin{eqnarray}
    V^\mathrm{dia}_r &=& -\frac{\partial \tilde{\eta}}{\partial r}
    + \sin^2\theta\frac{\partial\eta_\|}{\partial r}
    \nonumber \\
    V^\mathrm{dia}_\theta &=& \sin\theta\cos\theta\ \frac{\partial \eta_\|}{\partial r}.
    \label{13}
\end{eqnarray}
The meridional pumping velocity results from the rotationally induced
anisotropy of convective turbulence \citep{KN16}. Also the extra
diffusivity $\eta_\|$ along the rotation axis in addition to its
isotropic part $\eta$ results from rotational anisotropy.

The differential rotation and meridional flow of our model are taken from
the hydrodynamical model by \citet{KO11,KO12}. The hydrodynamical model
does not prescribe the eddy transport coefficients but defines them in
terms of the entropy gradient that is controlled by the (nonlinear) heat
transport equation. Recently \citet{Jea18} discussed the corresponding
closure in the convective turbulence theory. The hydrodynamical model
therefore supplies the diffusivity profiles also for our dynamo model.

The diffusivity value is close to $3\times
10^{12}$\,cm$^2$s$^{-1}$ in the bulk of the convection zone.
The diffusivity is however reduced by about four orders of magnitude in the
thin ($\sim$2\%$R_\sun$) near-bottom (overshoot) layer.
The diffusion inhomogeneity results in downward diamagnetic pumping that can be significant for the performance of solar dynamo models
\citep{KKT06,GG08,KC16}.
In our model, the pumping concentrates the
magnetic fields in the thin near-bottom layer of low diffusion, thus
enabling the coexistence of a weak radial field on the surface with kilo-Gauss
near-bottom toroidal fields. In this sense, the model realizes an
interface dynamo in spite of its distributed-type mathematical
formulation.

The equation for the quadrupolar poloidal field can be obtained from
Equation\,(\ref{12}) by the interchange of the parity indexes
$\mathrm{d}\leftrightarrow \mathrm{q}$. Nevertheless, the equation has
to be written explicitly,
\begin{eqnarray}
    \frac{\partial A^\mathrm{q}}{\partial t}
    &=& - \left(V^\mathrm{m}_r + V^\mathrm{dia}_r\right)\frac{\partial A^\mathrm{q}}{\partial r}
    - \left(V^\mathrm{m}_\theta + V^\mathrm{dia}_\theta\right)\frac{1}{r}
    \frac{\partial A^\mathrm{q}}{\partial \theta}
    \nonumber \\
    &+& \eta\frac{\partial^2A^\mathrm{q}}{\partial r^2}
    + \frac{\eta}{r^2}\sin\theta\frac{\partial}{\partial\theta}\frac{1}{\sin\theta}
    \frac{\partial A^\mathrm{q}}{\partial\theta} + \eta_\|\frac{\partial^2A^\mathrm{q}}{\partial z^2}
    \nonumber \\
    &+& r \alpha\ \frac{\sin^{n_\alpha + 1}\theta\ \phi_\alpha(r)}{1 + E_\mathrm{m}/E_0}
    \bigg( \cos\theta B^\mathrm{q}(r_\mathrm{i},\theta)
    \nonumber \\
    &+& \frac{\sigma}{\sqrt{2}}S^\mathrm{a}(t)\cos\theta B^\mathrm{q}
    + 0.1\sigma S^\mathrm{s}(t) B^\mathrm{d}(r_\mathrm{i},\theta)\bigg) ,
    \label{14}
\end{eqnarray}
for subsequent discussions.

Equations for the toroidal dipolar and toroidal quadrupolar fields include fields
of the same parity only. The equations repeat the toroidal field
equation of the model that does not separate fields of different parity
\citep[Equation\,(10) in][]{KN17} and we do not reproduce this equation here.

Our dynamo equations are fully symmetric about the interchange
$\mathrm{d}\leftrightarrow \mathrm{q}$. The symmetry is however broken
by the equatorial conditions
\begin{equation}
    B^\mathrm{d} = \frac{\partial A^\mathrm{d}}{\partial\theta}
    = \frac{\partial B^\mathrm{q}}{\partial\theta} = A^\mathrm{q} = 0 ,
    \label{15}
\end{equation}
which are used for solving the dynamo equations in one, say, northern
hemisphere only; the extension to the other hemisphere is easy.

The threshold value of the amplitude $\alpha^\mathrm{d}=0.158$\,m\,s$^{-1}$
of the $\alpha$-effect for the onset of the dynamo-instability to the dipolar fields is smaller than
$\alpha^\mathrm{q}=0.169$\,m\,s$^{-1}$ for the quadrupolar fields.
The dipolar mode is therefore dominant in our model. Observational gyrochronology shows that the solar rotation rate exceeds by about 10\% the rate of magnetic activity
termination for the solar twins \citep{MvS17}. We therefore put the
$\alpha$-parameter $\alpha = 0.174$\,m\,s$^{-1}$ about 10\% larger than its threshold value  in all computations to follow.

Dynamos are known to be self-sustained mechanisms that amplify
preexisting seed magnetic fields. Dynamo equations do not include the
sources of a magnetic field. This is not exactly so, however, in the NS
asymmetry problem. Whenever the last term - $S^\mathrm{s}(t)
B^\mathrm{d}$ - on the right-hand side of Equation\,(\ref{14}) was deleted,
the numerical solution converged to pure dipolar parity upon
sufficiently long time irrespective of the initial field parity.
Keeping this term finite always resulted in solutions of mixed parity.
The last term in Equation\,(\ref{14}) is, therefore, the source for
quadrupolar fields, and the mechanism for NS asymmetry in our model is
the quadrupolar field excitation by the dominant dipolar dynamo mode.

A disturbing feature of the model is that the correlation time of the
fluctuations is about one hundred times shorter than the oscillation
period of the quadrupolar fields. It may seem doubtful that short-term
fluctuations can produce quadrupolar oscillations with long-term coherence.
We proceed by showing that the doubts are misleading by considering an
illustrative example of regular oscillations excited by short-term random
forcing with zero mean.
\section{Illustrative example: randomly forced oscillator}
The simplest case of forced oscillations is governed by the equation
\begin{equation}
    \ddot{x} + \omega_0^2 x + 2\nu \dot{x} = f(t),
    \label{16}
\end{equation}
where $f$ is the driving force, and the friction coefficient $\nu$
($\nu < \omega_0$) is introduced to limit the oscillator's memory time,
which is infinite otherwise. The dots above the \lq displacement' $x$
mean the time derivatives.

The simplicity of Equation\,(\ref{16}) permits its solution for arbitrary forcing
function. The solution for the initial conditions of $\dot{x} = 0,\ x =
0$ at $t = 0$ reads
\begin{equation}
    x(t) = \omega^{-1}\int\limits_0^t \mathrm{e}^{-\nu t'}
    \sin(\omega t') f(t - t')\, \mathrm{d}t',
    \label{17}
\end{equation}
where $\omega = \sqrt{\omega_0^2 - \nu^2}$ is the oscillation frequency.

Let $f(t)$ be a statistically steady random function so that the
correlation
\begin{equation}
    F(t) = \langle f(T+t)f(T)\rangle
    \label{18}
\end{equation}
depends on the difference $t$ of the correlated function arguments.

Upon sufficiently long time $t \gg \nu^{-1}$, the oscillator forgets the
initial conditions and the upper limit in the integral Equation\,(\ref{17}) can be put to infinity.
The displacement $x(t)$ for this asymptotic solution is a statistically
steady random function of time as well with its correlation function
\begin{eqnarray}
    X(t) &=& \langle x(T+t) x(T)\rangle
    = \omega^{-2}\int\limits_0^\infty \int\limits_0^\infty
    \mathrm{e}^{-\nu(\tau + \tau')}
    \nonumber \\
    &\times& \sin(\omega\tau)\sin(\omega\tau')
    F(t -\tau + \tau')\,\mathrm{d}\tau\,\mathrm{d}\tau'.
    \label{19}
\end{eqnarray}

The case of the correlation time $\tau_\mathrm{c}$ of random forces being short compared with the oscillation period can be represented by the delta-correlated in time random forcing
\begin{equation}
    F(t) = \langle f^2\rangle\tau_\mathrm{c}\delta(t) .
    \label{20}
\end{equation}
Substitution of this equation into Equation\,(\ref{19}) leads to the expression for the displacement correlation function
\begin{equation}
    X(t) = \frac{1}{2}A^2 C(t)
    \label{21}
\end{equation}
in terms of the {\sl rms} amplitude
\begin{equation}
    A = \frac{1}{\omega_0}\sqrt{\frac{\langle f^2\rangle \tau_\mathrm{c}}{2\nu}}
    \label{22}
\end{equation}
and normalized correlation function
\begin{equation}
    C(t) = \mathrm{e}^{-\nu\mid t\mid}\left[ \cos (\omega t)
    + \frac{\nu}{\omega}\sin(\omega\mid t\mid)\right].
    \label{23}
\end{equation}

In spite of the driving force being on average zero, the amplitude of the forced oscillation is finite. A probable explanation is as follows. In the course of one correlation time, the random force increases or decreases the oscillator velocity by a small amount $\delta v \sim \langle f^2\rangle^{1/2}\tau_\mathrm{c}$ with equal probability. The velocity squared, however, increases on average by an amount of $\delta v^2/2$ in one forcing event. The energy growth of the forced {\em linear} oscillator saturates due to finite friction only.

\begin{figure}
\includegraphics[width=\columnwidth]{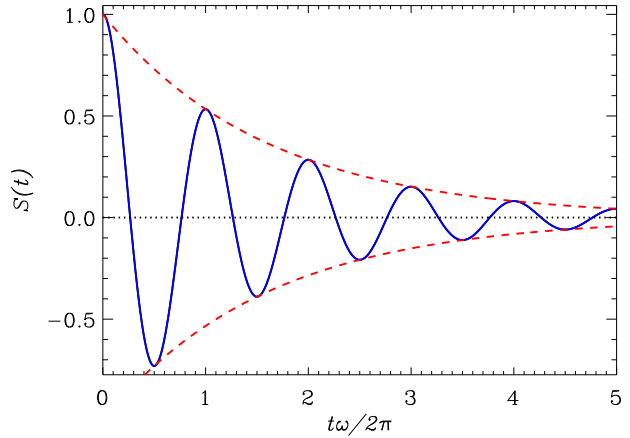}
\caption{Correlation function of Equation\,(\ref{23}) for $\nu = 0.1\omega$
        (full line). The dashed lines show the $\pm\exp(-\nu t)$ envelope.}
         \label{f1}
\end{figure}

The average amplitude (\ref{22}) of the asymptotic oscillations is steady. The correlation function of Figure\,\ref{f1} shows decreasing oscillations however. The correlation function of strictly periodic oscillation is periodic as well. The decreasing amplitude in Figure\,\ref{f1} results from phase wandering of the randomly forced oscillation. Equation\,(\ref{23}) shows that the coherence time of the oscillations $1/\nu$ is also controlled by the friction. The coherence time does not depend on the correlation time of random forcing. The short correlation time therefore does not exclude long-term coherence of the randomly forced oscillation.

\begin{figure*}[t!]
\center{\includegraphics[width = 15 truecm]{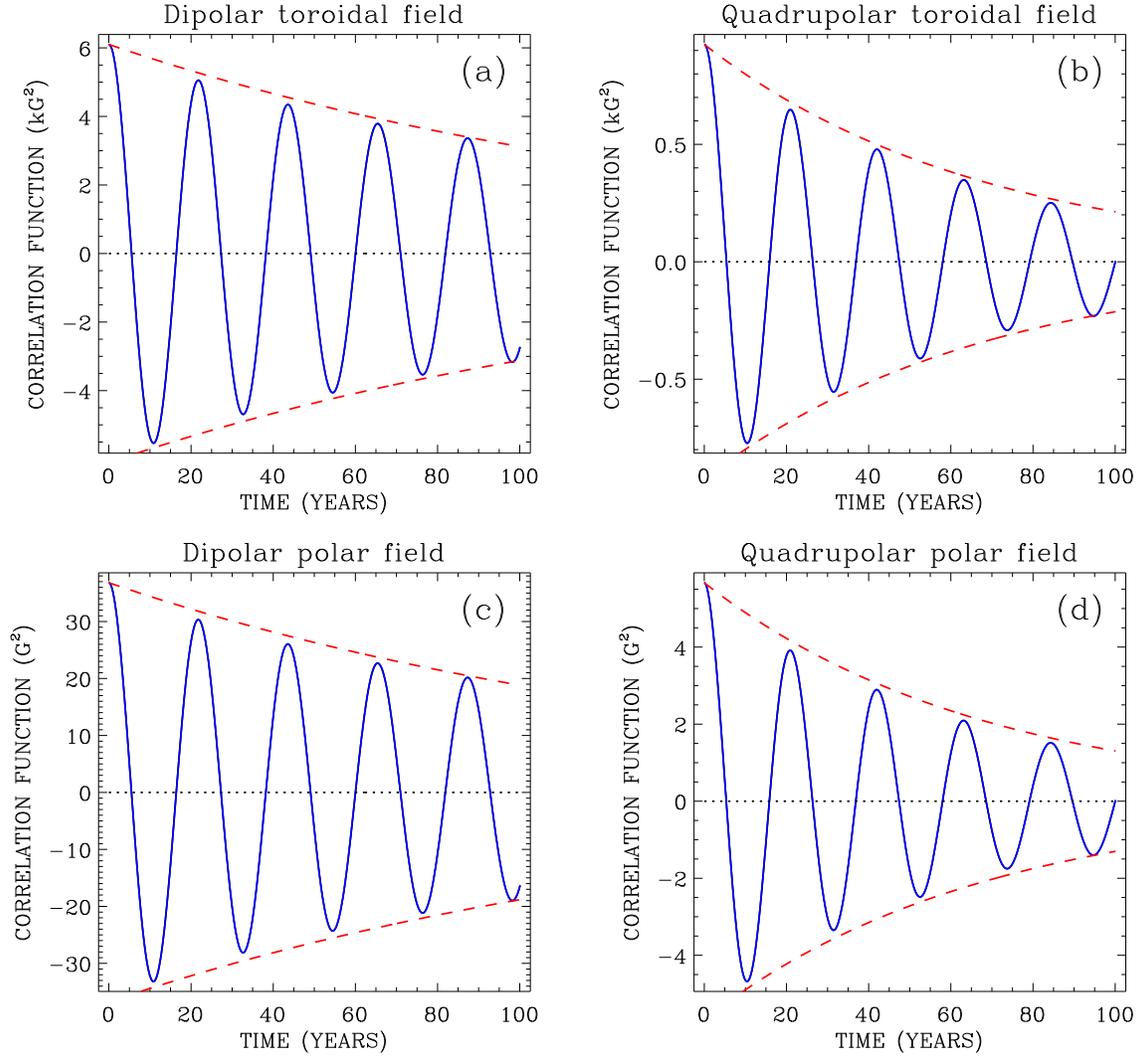}}
\caption{Correlation functions of Equation\,(\ref{24}) for the dipolar
        (a) and quadrupolar (b) bottom toroidal fields and the correlation functions  for the polar surface radial fields of dipolar (c) and quadrupolar (d) parity. The dashed lines show the approximate exponential envelopes
        $\pm A^2 \exp(-t/t_\mathrm{c})/2$.
        }
        \label{f2}
\end{figure*}

The correlation function of Equation\,(\ref{19}) turns out to be an informative characteristic of random oscillation. The function value at zero argument defines the mean oscillation amplitude and the function profile defines the coherence time. We therefore use correlation functions for  oscillatory dynamo modes in the following discussion of the model results.
\section{Results and discussion}
Computations of this paper started with a preliminary run of about one thousand years physical time. The aim of this run was to arrive at a state statistically independent of the initial condition. This state was then used as the initial condition for the main run of eleven thousand years (about ten thousand dynamo-cycles). The main run provided the statistical base for the results.
\subsection{Origin and coherence time of NS asymmetry}
Figure\,\ref{f2} shows the correlation functions of dipolar and quadrupolar toroidal fields at the base of the convection zone and latitude of 15$^\circ$ close to the latitude where the toroidal field attains its maximum strengths in the dynamo cycles. The correlation functions were computed with numerical time-integration
\begin{equation}
    C_\mathrm{tor}^\mathrm{p}(t) = \frac{1}{T-t}\int\limits_{0}^{T-t}
    B^\mathrm{p}(r_\mathrm{i},t'+t)
    B^\mathrm{p}(r_\mathrm{i},t')\,\mathrm{d}t' ,
    \label{24}
\end{equation}
where the upper parity index \lq$\mathrm{p}$' is \lq$\mathrm{d}$' or \lq$\mathrm{q}$' for the dipolar or quadrupolar fields of Equations\,(\ref{6}) and (\ref{7}), respectively, and $T = 11000$\,years of the main run. Figure\,\ref{f2} also shows the correlation functions for the surface polar fields of dipolar and quadrupolar parity.

\begin{figure*}[t!]
\center{\includegraphics[width = 15 truecm]{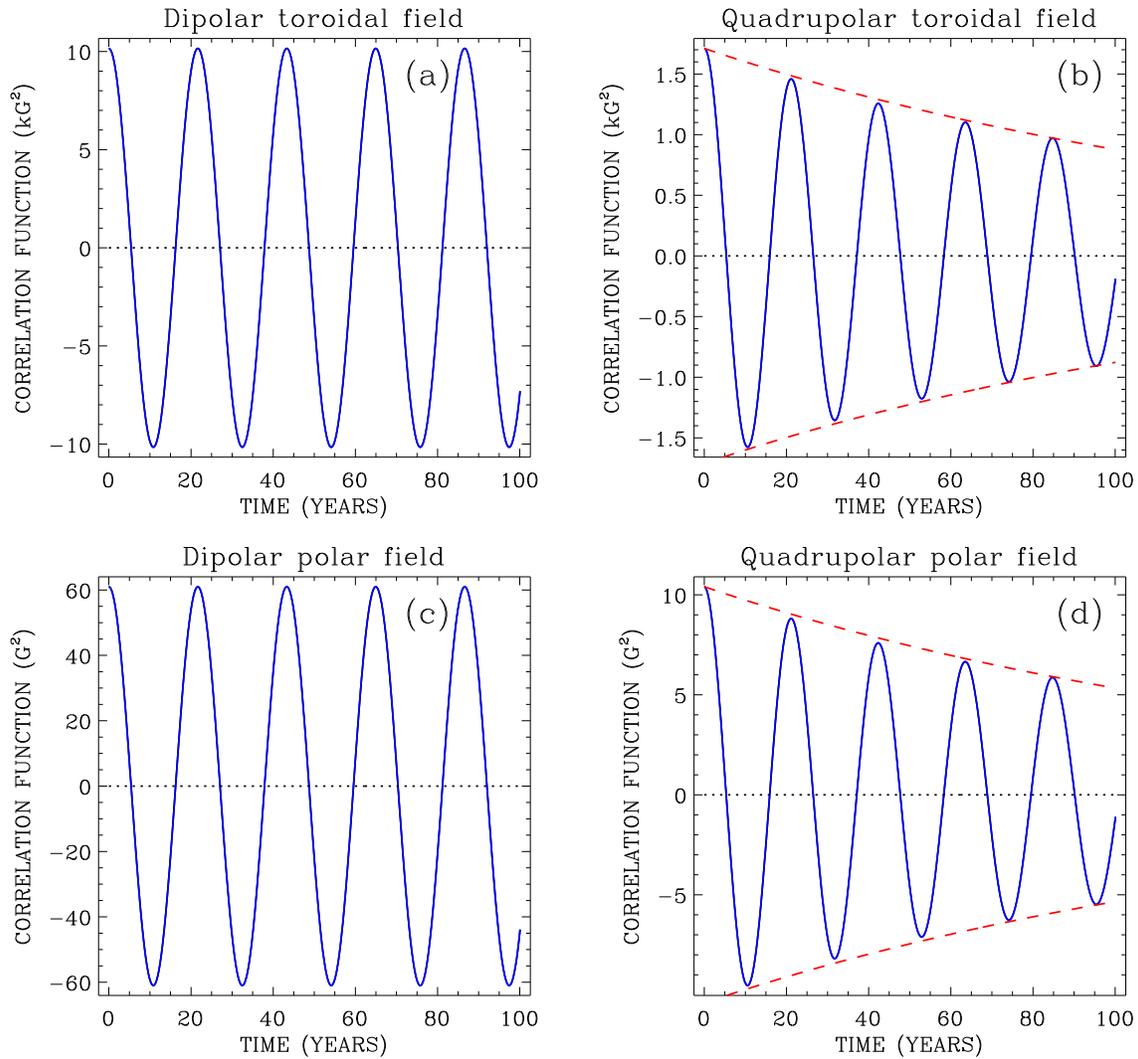}}
\caption{The same as in Figure\,\ref{f2} but for computations with a reduced model
         that keeps fluctuations in the last term of Equation\,(\ref{14}) only and replaces total magnetic energy $E_\mathrm{m}$ by the energy of the dipolar magnetic field in the $\alpha$-effect of Equation\,(\ref{3}).
         }
        \label{f3}
\end{figure*}

The correlation functions of Figure\,\ref{f2} are similar to the function of Figure\,\ref{f1} of  randomly forced oscillation. The wavy sign-changing profiles of this functions are evident for long-term coherence of the quasi-periodic oscillations. Envelopes of the functions are closely approximated by the exponential profiles
\begin{equation}
    C_\mathrm{env}(t) = \pm A^2 \exp(-t/t_\mathrm{c})/2,
    \label{25}
\end{equation}
where $A$ is the {\sl rms} amplitude the oscillations and $t_\mathrm{c}$ is the coherence time. According to Figure\,\ref{f2}, the amplitude of the bottom quadrupolar field of about 1.36\,kG is about 2.6 times smaller compared to the dipolar toroidal field amplitude of 3.49\,kG. The amplitude of the quadrupolar polar field of 3.37\,G is smaller than the dipolar field amplitude of 8.58\,G in about the same proportion. The resulting NS asymmetry is therefore small as it is on the Sun \citep{NCP14}.

The envelope lines for quadrupolar field correlations in Figure\,\ref{f2} show almost equal coherence time $t_\mathrm{c}^\mathrm{q} \simeq 68$\,yr for the quadrupolar toroidal and surface polar fields. This is about 6.5 times longer that the mean period $\overline{P}^\mathrm{q}_\mathrm{cyc} = 10.4$\,yr of the quadrupolar (energy) cycle. The mean cycle period of dominant dipolar dynamo mode $\overline{P}^\mathrm{d}_\mathrm{cyc} = 10.9$\,yr is a bit longer. The quadrupolar oscillations randomly forced by the dominant dipolar dynamo mode have long-term coherence. The close approximation of the correlation function peaks in Figure\,\ref{f2} by the exponential profiles of Equation\,(\ref{25}) shows that the bottom toroidal and the surface polar fields vary smoothly with time. The small overestimation of the peaks by the exponential envelopes can be attributed to minor irregular variations in the fields on the short time-scale of the $\alpha$-effect fluctuations.

The finite coherence time $t_\mathrm{c}^\mathrm{d} = 149$\,yr of the dominant dipolar mode in Figure\,\ref{f2} is a consequence of phase wandering caused by fluctuations in the $\alpha$-effect. In order to test this statement, computations were repeated with fluctuations of the $\alpha$-effect in Equations\,(\ref{12}) and (\ref{14}) switched off everywhere except for the last term of Equation\,(\ref{14}) and the total magnetic energy $E_\mathrm{m}$ in Equation\,(\ref{3}) replaced by the energy of the dipolar field alone. In this case, the dipolar mode decouples from the quadrupolar mode. The results of this \lq reduced' model are shown in Figure\,\ref{f3}. As expected, the dipolar mode is strictly periodic in this case. The quadrupolar quasi-periodic oscillations are still present with their coherence time increased to $t^\mathrm{q}_\mathrm{c} \simeq 150$\,yr. The field amplitudes also increased due to weaker magnetic quenching of the $\alpha$-effect in the reduced model.

Several other computations were done to confirm that the dynamo eventually converges to pure dipolar parity whenever the last term in Equation\,(\ref{14}) is omitted. Alternatively, quadrupolar oscillations with long-term coherence were always present when this term was retained irrespective of whether fluctuations in other dynamo parameters were retained or not. This confirms the quadrupolar mode excitation by the dominant dipolar mode via the equator-symmetric short-term fluctuations in the $\alpha$-effect as the viable dynamo mechanism for the NS magnetic asymmetry.

The results to follow correspond to complete model computations.
\subsection{Phase locking}
\begin{figure}
\includegraphics[width=\columnwidth]{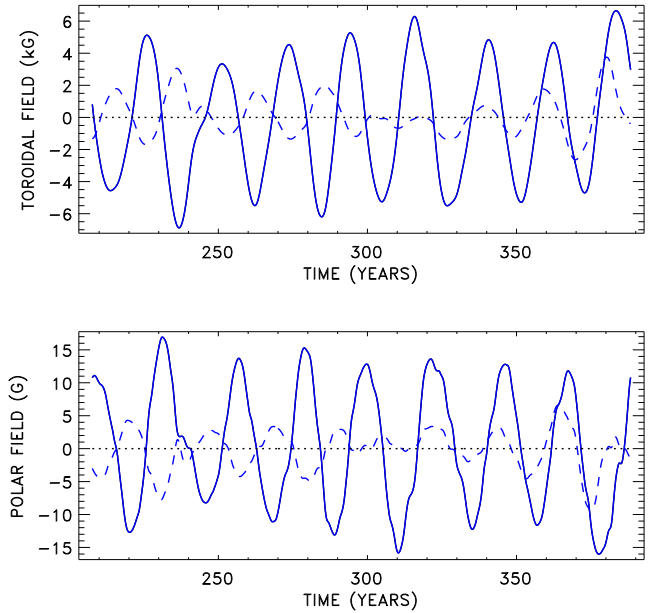}
\caption{{\sl Top panel}: dipolar (full line) and quadrupolar (dashed) toroidal
         field at the 15$^\circ$ latitude of the bottom of the convection zone as the function of the run time. {\sl Bottom panel}: the plot of dipolar (full line) and quadrupolar (dashed) northern polar fields. This computation fragment illustrates the phase-locking phenomenon.
         }
         \label{f4}
\end{figure}

\begin{figure}
\includegraphics[width=\columnwidth]{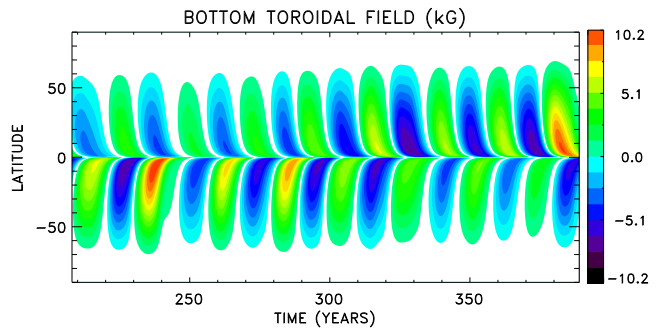}
\caption{Time-latitude diagram of the bottom toroidal field for the same computation fragment
         as Figure\,\ref{f4}.
         }
         \label{f5}
\end{figure}

The oscillations of dipolar and quadrupolar fields tend to proceed almost in phase or in antiphase with relatively fast transitions between these two cases. Figures\,\ref{f4} and \ref{f5} give a characteristic example of this tendency. The fields of different parity are almost in antiphase until the run time of 300 yr and change to almost in-phase oscillation after 320 yr.
The in-phase and antiphase oscillations differ by the sense of the NS asymmetry: the fields are predominantly of the same sign in the northern hemisphere and of mainly opposite sign in the southern hemisphere for in-phase oscillations and vice versa for the antiphase oscillations (Figure\,\ref{f5}).

A quantitative characteristic of this phase-locking phenomenon can be the correlation
\begin{equation}
    C_\mathrm{PL} = \langle \left(\mid B^\mathrm{d}\mid
    - \langle \mid B^\mathrm{d}\mid \rangle\right)
    \left( \mid B^\mathrm{q}\mid
    - \langle \mid B^\mathrm{q}\mid \rangle \right)\rangle ,
    \label{26}
\end{equation}
where $B^\mathrm{d}$ and $B^\mathrm{q}$ are magnetic fields of dipolar and quadrupolar parity at some location in the convection zone and angular brackets mean averaging over time of the run.
The correlations of Equation\,(\ref{26}) have positive values for the bottom toroidal field at the 15$^\circ$ latitude ($C_\mathrm{PL}$=0.54\,kG$^2$) and for the surface polar field ($C_\mathrm{PL}$=3.28\,G$^2$). The positive values mean that the dipolar and quadrupolar fields tend to be above or below their mean strength simultaneously, i.e. they tend to oscillate in phase or in antiphase. The probable reason for this synchronization is that random forcing of the quadrupolar dynamo mode is modulated by oscillations of the dipolar toroidal field.

The phase locking may be the explanation for the observation that \lq\lq two hemispheres never get very far out of phase with each other" \citep{H15}.
\subsection{Large asymmetry in Grand minima}
\begin{figure}
\includegraphics[width=\columnwidth]{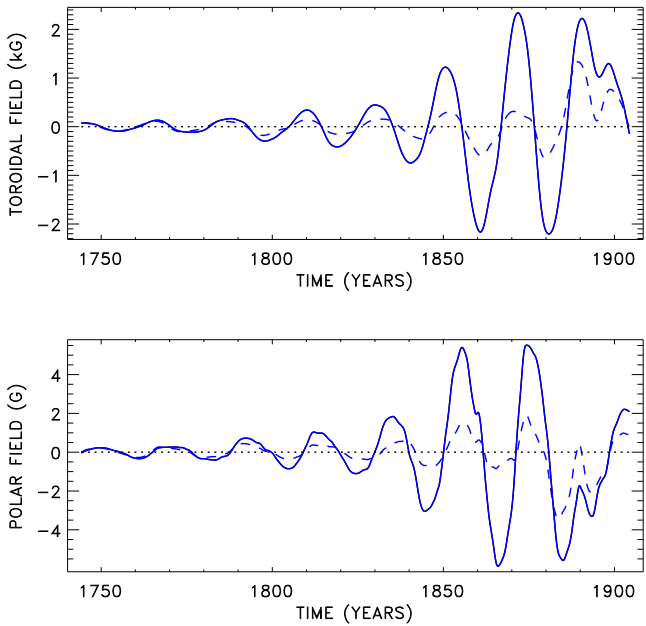}
\caption{The top panel shows the dipolar (full line) and quadrupolar (dashed)
         toroidal field at the 15$^\circ$ latitude of the bottom of the convection zone at the end of an epoch of weak cycle amplitude in the dynamo simulation. The bottom panel shows the dipolar (full line) and quadrupolar (dashed) northern polar fields.
         }
         \label{f6}
\end{figure}

\begin{figure}
\includegraphics[width=\columnwidth]{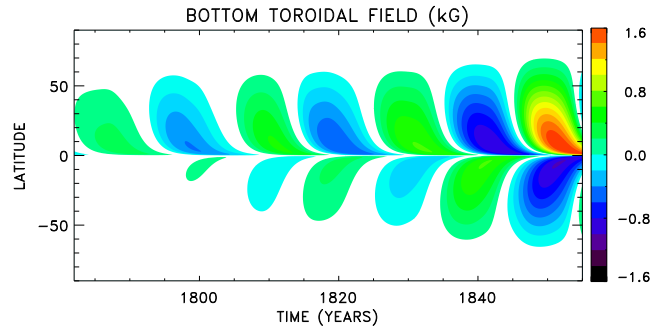}
\caption{The butterfly diagram of the bottom toroidal field for a part of
         computation fragment of Figure\,\ref{f6}.
         }
         \label{f7}
\end{figure}

Dynamo models with fluctuating parameters are known to show transient epochs of strongly reduced field strength modeling the Grand minima of solar activity \citep{Cha20}. The epochs of weak magnetic fields usually - though not always - show up-normal NS asymmetry. A characteristic example is shown in Figures\,\ref{f6} and \ref{f7}.

The increased asymmetry in the simulated Grand minima can be interpreted as follows. The dominant dynamo mode in our model is the dipolar one. This is because the critical value $\alpha^\mathrm{q}=0.169$\,m\,s$^{-1}$ of the $\alpha$-parameter for the quadrupolar mode excitation is larger than its value for the dipolar mode. In computations without fluctuations in $\alpha$, the field eventually relaxes to dipolar parity. However, if the computations started from weak initial field of mixed parity, fields of either parity initially grow. This is because the 10\% supercritical value of $\alpha = 0.174$\,m\,s$^{-1}$ in the computations is larger than $\alpha^\mathrm{q}$. As magnetic energy grows, the magnetic quenching of Equation\,(\ref{3}) reduces the \lq effective' $\alpha$-value below $\alpha^\mathrm{q}$ and the quadrupolar mode changes to decay. If fluctuations in dynamo parameters drive the dynamo into a deep minimum, the magnetic quenching becomes inefficient. The fields of either parity start growing afterwards so that amplitudes of both parities can be comparable at the end of the simulated Grand minimum.

Sunspot data of Paris Observatory show that almost all sunspots of the end of the Maunder minimum emerged in the southern hemisphere \citep{RNR93}. So large hemispheric asymmetry can be interpreted in terms of the dynamo theory as almost equal amplitudes of dipolar and quadrupolar fields \citep{SNR94}.
\subsection{Is the NS asymmetry predictable?}
The $\alpha$-effect of the $\alpha\Omega$ dynamo converts the toroidal magnetic field into a poloidal one. The dynamo loop is closed by the $\Omega$-effect of poloidal-to-toroidal field conversion by differential rotation. In contrast to the $\alpha$-effect, the solar differential rotation does not include any substantial randomness. A functional relation can therefore be expected between the polar (poloidal) field of minimum activity epochs and the strengths of the following activity cycles \citep{Sea78}. A tight correlation of this type is confirmed observationally \citep{MJea13,HU16} and reproduced by dynamo models \citep{Cha20,Kea21}. The polar field strength is recognized as a robust precursor for solar cycles' strength \citep{Pet20,Nan21}.

The characteristic time $R_\odot^2/\eta$ of inter-hemispheric diffusive coupling exceeds the solar cycle period for the observations-based estimation $\eta \simeq 3\times 10^{12}$\,cm$^2$s$^{-1}$ of the eddy diffusivity \citep{CS16}. The polar field asymmetry can therefore be expected to be a precursor for sunspot asymmetry.

\begin{figure}
\includegraphics[width=\columnwidth]{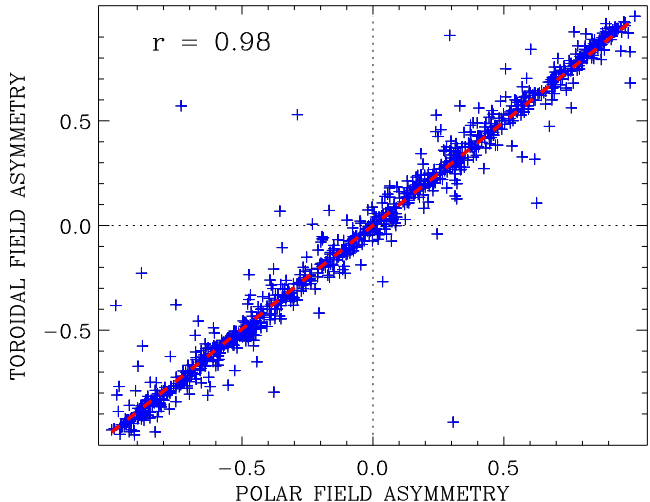}
\caption{Scatterplot of the computed cycles positions on the plane of the polar
         field asymmetry of Equation\,(\ref{27}) at the cycle's minima and the toroidal field asymmetry Equation\,(\ref{28}) at the maxima of the following cycles. The dashed line shows the best linear fit.
         }
         \label{f8}
\end{figure}

Figure\,\ref{f8} shows the computed cycle positions on the plane of the polar field asymmetry
\begin{equation}
    A_\mathrm{pol} = \frac{B_r^2(90^\circ) - B_r^2(-90^\circ)}
    {B_r^2(90^\circ) + B_r^2(-90^\circ)}
    \label{27}
\end{equation}
at the cycles' minima and the toroidal field asymmetry
\begin{equation}
    A_\mathrm{tor} = \frac{B^2(15^\circ) - B^2(-15^\circ)}
    {B^2(15^\circ) + B^2(-15^\circ)}
    \label{28}
\end{equation}
at the following cycles' maxima. In these equations, $B_r$ is the surface radial field, $B$ is the bottom toroidal field, and the field argument mean the latitude where the field values were taken. The computed cycle minima were defined as the instants of the dipolar bottom toroidal field reversals at the latitude of $15^\circ$ and the cycle maxima - as the instants of maximum strength of this field.

The correlation coefficient for NS asymmetries of Figure\,\ref{f8} is high, $r = 0.98$. The correlation agrees with the dynamo model by \citet{NLC19}. Though the correlation of Figure\,\ref{f8} is not as tight as the correlation between the polar and toroidal field strengths \citep[see Figure\,4 in][]{KMN18}, it suggests that the polar field asymmetry at the solar minima can serve as precursor for the asymmetry of the following sunspot cycle.

\begin{figure}
\includegraphics[width=\columnwidth]{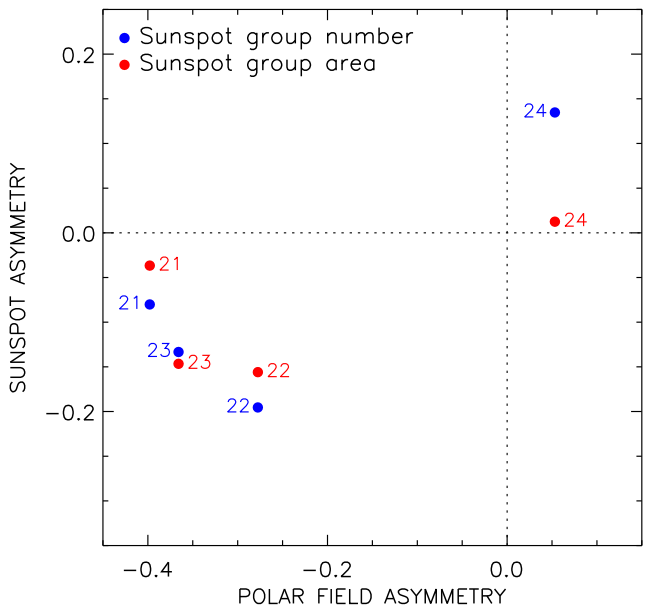}
\caption{Positions of solar cycles 21 to 24 on the plane of polar flux asymmetry at
         the minima before these cycles and the NS asymmetry of the cycles in terms of the total sunspot area and the sunspot group number.
         }
         \label{f9}
\end{figure}

Whether this model-based correlation agrees with observations is not clear. \citet{MJea13} analyzed the correlation of the polar field proxies and the strength of following sunspot cycles separately for the northern and southern solar hemispheres. Their Figure\,3 shows that those hemisphere that had larger polar flux prior to the onset of the cycles 19 to 22 had a larger amplitude of sunspot area in the following cycles. The earlier cycles 15, 17, and 18 showed the opposite relation, however.

Figure\,\ref{f9} shows the relation between the polar field asymmetry with the sunspot asymmetry for the solar cycles 21 to 24 for which measurements of the polar field are available. The polar field strength was estimated from synoptic maps of the Mount Wilson Observatory\footnote{\url{ ftp://howard.astro.ucla.edu/pub/obs/synoptic_charts/fits/}}. For each synoptic map, the magnetic fields at the latitudes between 55 degrees and the poles were averaged in both hemispheres and then smoothed by a 13 month running mean to filter-out the yearly geometric projection effects. After that, the maximal strengths of the northern and southern polar fields at the sunspot activity minima were determined. The sunspot data of Figure\,\ref{f9} were taken from the USAF/NOAA data files\footnote{\url{http://solarcyclescience.com/activeregions.html}}. For each sunspot group, the data in the sunspot area maxima were chosen. The sunspot groups were sorted by the solar cycles according to the method used by \citet{MNL14}. Then the total number and area of sunspot groups in the solar cycles were calculated. The relative asymmetry $(N^2 - S^2)/(N^2 + S^2)$ was used in this Figure, where N and S are the maximum strengths of the northern and southern polar fields or total sunspot group number or area in the northern and southern hemispheres, respectively.

The low statistics of Figure\,\ref{f9} does not permit its quantitative analysis. The polar field asymmetry prior the four cycles of this Figure, however, coincides in its sense with the sunspot asymmetry of the following cycles.
\section{Conclusions}
The long-term component in the hemispheric asymmetry of solar activity can be caused by  short-term fluctuations in the BL dynamo mechanism. This conclusion follows from the dynamo model with a fluctuating $\alpha$-effect.

The dynamo model separates the equations for equator-symmetric (quadrupolar) and equator-antisymmetric (dipolar) magnetic fields. The equations are linked by the equator-symmetric fluctuations in the BL-type $\alpha$-effect. The model shows NS asymmetry when the fields of different equatorial parity are excited simultaneously. The dominant dipolar dynamo mode is excited irrespectively of the fluctuations in the $\alpha$-effect. The quadrupolar mode on the contrary decays whenever the equator-symmetric fluctuations in $\alpha$ are switched off. This behavior permits us to propose the excitation of quadrupolar oscillations by a dominant dipolar dynamo mode mediated by the equator-symmetric fluctuations in the $\alpha$-effect as the physical mechanism for NS asymmetry of solar activity.

The characteristic time of fluctuations in the dynamo model is of the order of one solar rotation period. The resulting NS asymmetry, however, varies on the time scale of several 11-year cycles. An analytically solvable example of an oscillator excited by short-term random forces shows that long-term coherence of randomly forced oscillations is not an exceptional property of oscillating dynamos.

The dynamo modeling shows the phase-locking phenomenon: the quasi-periodic quadrupolar field varies almost in phase or in antiphase with the dipolar field with relatively short irregular transitions between these two states. This means that magnetic field oscillations in different hemispheres show irregular wandering of amplitude and phase but their going far out of phase are relatively short and seldom. The variations of NS asymmetry in the model are not periodic.

The model also shows increased NS asymmetry in the (Grand minima) epochs of weak magnetic fields. This behavior can be explained by weaker magnetic quenching of the $\alpha$-effect during the simulated Grand minima.

The model computations suggest the NS asymmetry of polar magnetic fields in the activity minima as a precursor of NS asymmetry of integral characteristics of sunspot activity in the following solar cycles.
\bl

This work was financially supported by the Ministry of Science and High
Education of the Russian Federation and by the Russian Foundations for Basic
Research (Project 19-52-45002\_Ind). This study includes data from the
synoptic program at the 150-Foot Solar Tower of the Mt. Wilson Observatory. The
Mt. Wilson 150-Foot Solar Tower is operated by UCLA, with funding from NASA, ONR,
and NSF, under agreement with the Mt. Wilson Institute.
The authors are thankful to Ms Jennifer Sutton for language editing.
\bibliography{Paper}{}
\bibliographystyle{aasjournal}
\end{document}